\documentstyle[12pt]{article}
\begin{document}
\title{ Dirac confinement of 
a heavy quark-light 
quark system $(Q,\bar{q})$ in high orbital 
angular momentum states. 
}
\author{
M. A. Avila \thanks{Electronic address: manuel@servm.fc.uaem.mx}
\\
  Facultad de Ciencias, UAEM, \\
      Cuernavaca 62210, Morelos, Mexico.\\  }

\date{}
\maketitle

\begin{abstract}

The Regge behaviour of the solutions of a  
Dirac hamiltonian describing a heavy quark-light quark
system in high orbital angular momentum states is
analyzed.
It is found that the solutions of a scalar confining potential
are physically admissible 
while those of a vector confining potential
are not. 
It is concluded that
with a Dirac hamiltonian a 
scalar confining potential
is preferred over a 
vector confining potential
for any value of the orbital angular
momentum. 
\end{abstract}

\hskip1.0cm

\noindent{PACS number(s) : 12.39.-x; \,
12.39.Pn; \, 12.39.-t}

\newpage

\normalsize

Recently 
there has been given a discussion about the 
nature of the confinement potential in a heavy quark - light quark
$(Q,\bar{q})$ system \cite{olsson}-\cite{avila}.
In Ref. \cite{olsson} an analysis was 
performed 
from a phenomenological point 
of view 
using diverse techniques. These were  
the calculation of the Isgur-Wise 
(IW) function \cite{hqet} as predicted by both
the  
Dirac and the no-pair equation, the study of 
the classical returning points for $s$-waves
(all of the above for 
low orbital angular momentum states ),
and the study of the Regge behavior 
of the no-pair equation describing
the $(Q,\bar{q})$ system    
in high orbital angular momentum states. 
In this work the authors  
found 
that with a Dirac-like equation 
only a Lorentz scalar confinement 
accounts for the unphysical 
phenomenon of the mixing
of negative with positive energy states
also called the Klein paradox,   
while with the no-pair variant of 
the Dirac equation only a Lorentz vector confinement 
potentials leads to a normal
Regge behaviour. Concerning to the 
calculation
of the IW function
they found that 
the no-pair equation
predicts a value
of its slope at
zero recoil point 
in better agreement with the 
heavy-light data than the Dirac equation.
The authors of \cite{olsson}
conclude arguing
against scalar confinement.

It is worth it to stress at this 
stage two 
points about the work of Ref. \cite{olsson}
that eventually modify 
the results.
The first one is that 
the analysis of Regge behavior was done  
partially, since only the 
the no-pair equation was considered, while   
the Dirac-like equation was not.
The other is that 
the relativistic
corrections to the Dirac hamiltonian
were not included.

On the other hand, in Ref. \cite{avila}
a $(Q,\bar{q})$ system 
was considered   
as described by a Dirac-like 
hamiltonian containing all the 
relativistic corrections.
By assuming that this system
is in low orbital
angular momentum states  
it was shown that only a 
scalar confinement 
leads to a finite norm
of the wave function.
Also, it was checked that only this 
potential accounts 
for the Klein paradox.
In \cite{avila}    
the slope of the IW function at 
zero recoil point including relativistic
corrections was also calculated
as predicted by  
scalar confinement.  
The value found for this 
quantity agrees very
well with heavy-light data. 
In Ref. \cite{avila} 
it was concluded that 
scalar confinement is favored
over vector confinement 
when the $(Q,\bar{q})$ meson is in low
orbital angular momentum states and it is 
described by a covariant 
Dirac equation.

On the basis of the above 
remarks, the purpose of this 
work is then  
to investigate the nature of the
confinement when the $(Q,\bar{q})$ system 
is described by a Dirac equation in
the Regge limit of 
high orbital angular momentum states.

In order to do the 
above we start with a simple model
where we are neglecting
the relativistic corrections to
the hamiltonian. Consequently,  
the hamiltonian for the c.m. system is 
\cite{olsson}, \cite{solo}

\begin{equation}
 \biggl[ {\bf \alpha}  \cdot 
{\bf p} \,\,+ \,\,
m \, \beta \,\, +\,\, U(r) \,\, + 
\,\, \beta S(r) \,+\, V(r) \biggr]
\psi   =  E\, \psi ,
\end{equation}

\noindent where $m$ ($M_Q$) is the light 
(heavy) quark 
mass, ${\bf p}$ is the momentum of the light quark, 
$\,U(r)\,=\,-\xi/r\,$ is a color Coulomb-like
potential, $\,S(r)\,=\,\kappa_s\,r\,$
and $\,V(r)\,=\,\kappa_v\,r\,$ are linear
increasing Lorentz scalar and vector potentials,
respectively.

The non-perturbative potentials
$S$ and $V$ in Eq. (1)
are dynamically responsible for the confinement
of the light quark, while the Lorentz vector 
potential $U$
describes the perturbative color interaction
between the quarks. 

If one writes (1) 
as a matrix equation  
then

\begin{eqnarray}
\left( 
\matrix{
m\,+\,S\,+\,U\,+\,V\,-E &  {\bf \sigma} \cdot {\bf p}
\cr
 {\bf \sigma} \cdot {\bf p}  & -m\,-\,S\,+\,U\,+\,V\,-E
}
\right)
\left(
\matrix{
G  \,\cr 
i {\bf \sigma}\cdot{ \bf \hat r} F \,
}
\right)\, \chi^m_\kappa\,=\left(
\matrix{
0 \cr 
0
}
\right).
\end{eqnarray}      

By using the identities 

\begin{equation}
{\bf \sigma} \cdot {\bf L}\,\chi^m_\kappa\,=\,-(1\,+\,\kappa)
\chi^m_\kappa ,
\end{equation}

\begin{eqnarray}
{\bf \sigma} \cdot {\bf p}\,\chi^m_\kappa \,=\,
i\, {\bf \sigma} \cdot \,{ \bf \hat r}
\Bigl(-\frac{d}{dr}\,+\,\frac{\kappa+1}{r}
\Bigr) \chi^m_\kappa,
\end{eqnarray}

\noindent where ${\bf L}\,=\,{\bf r} \times {\bf p}$      
is the angular momentum and 
\begin{equation}
\kappa \,
= \,
\cases{  -(l\,+\,1) & j=l+1/2 
\cr
\,\,\,\,\,\,\,\,\,\,\,\,\,\, l & j=l-1/2,\cr}
\end{equation}
\noindent Eq. (2) leads
to the following system of two 
coupled linear differential
equations

\begin{equation}                   
\frac{dG}{dr}\,=\,-\frac{\kappa+1}{r}\,G
\,-\, \biggl[
E\,+m\,+\,S\, -\,\bigl(
U\,+\,V
\bigr)
\biggr]\,F,
\end{equation}

\begin{equation}
\frac{dF}{dr}\,=\, \biggl[
\,E\,-\,\bigl(
m\,+\,S\,+\,U\,+\,V
\bigr)
\biggr]\,G\,+\,\frac{\kappa-1}{r}\,F.
\end{equation}

In order to analyze the Regge behavior
of the solutions of these equations
we make the following two approximations
for large values of the orbital angular 
momentum ($l\,\gg\,1$),

\begin{equation}
E\, =\,\sqrt{p_r^2\,+\, \frac{l(l+1)}{r^2}\,+m^2}\,\sim \,l/r,
\end{equation}

\begin{equation}
\mid \kappa \mid \,\pm\,1 \sim \, \mid \kappa\mid \,\sim \, l. 
\end{equation}

In order to assure the validity of 
Eq. (8) for any value of
$r$ and not only near the turning points, 
in this work it is assumed  
that the Regge objects under 
study have finite size of radius $R$.
Consequently from 
the Uncertainty Principle 
in the way $1\,\leq\,p_rR\,<\,l$, it
follows that for $r\,\leq \,R$

\begin{equation}
p_r^2+\frac{l(l+1)}{r^2}\,\simeq
\frac{(p_r r)^2+l^2}{r^2}\leq
\frac{(p_r R)^2+l^2}{r^2}\simeq \frac{l^2}{r^2}. 
\end{equation}

\noindent With which Eq. (8) 
holds for any value of $r$.

Before of finding the Regge 
solutions
let us find first the Regge slopes 
\footnote{ 
As it is well known for large orbital
angular momentum, the Regge trajectories
become linear for linear confinement
and have slopes whose values depend
on the Lorentz nature of the confinement \cite{olsson}. }
as predicted by Eqs. (6)-(9).
In order to find these  
we are assuming 
$\mid{\kappa}\mid \gg 1$ and nearly circular orbits
for the light quark.
From (6)-(9) it follows that

\begin{equation}
-\frac{F}{G}=\frac{V-E+S}{  \frac{\kappa}{r}}
=\frac{\frac{\kappa}{r}}{V-E-S}, 
\end{equation}

\noindent which is equivalent to

\begin{equation}
(E-V)^2-S^2=\frac{\kappa^2}{r^2}. 
\end{equation}

\noindent At the lowest energy state, the 
energy satisfies $\frac{\partial E}{\partial r}|_{\kappa}=0$
which implies

\begin{equation}
(E-V)V^\prime + S S^\prime = 
\frac{\kappa^2}{r^3}.
\end{equation}

From (12) and (13) we find the general 
expression for the Regge slope 

\begin{equation}
\alpha^\prime = \frac{\mid \kappa \mid}{E^2}=
\frac{\sqrt{ 2 \kappa_v\,\left(\kappa_v+ 
\sqrt{\kappa_v^2+8\kappa_s^2}\,\,\right)+4\kappa_s^2} }
{5\kappa_v^2+3\kappa_v \sqrt{\kappa_v^2+8\kappa_s^2}
+4\kappa_s^2}. 
\end{equation}
 
\noindent For the most interesting
particular cases, Eq. (14)
yields the following results

\hskip0.5cm - Scalar confinement: $V=0\,\,$
and $\,\,S=\kappa_s r$
\begin{equation}
\alpha^\prime=\frac{1}{2\kappa_s}
\end{equation}

\hskip0.5cm - Vector confinement: $V=\kappa_v r\,\,$
and $\,\,S=0$
\begin{equation}
\alpha^\prime=\frac{1}{4\kappa_v}
\end{equation}

\hskip0.5cm - Vector and Scalar confinement with same strength: 
$V=S=ar\,\,$

\begin{equation}
\alpha^\prime=\frac{1}{3 \sqrt{3} a}.
\end{equation}
 
We must note that these Regge 
slopes coming from the Dirac equation
(1) are the same than those previously
found in Ref. \cite{universal}
which were obtained from a general confinement
model for a $(Q,\bar{q})$ system.  
For the description of this system in \cite{universal}  
it was employed 
a Klein-Gordon equation 
whith a confinement  
potential
being a linear combination
of scalar and 
time-component vector potentials.  
The coincidence between the values 
of the slopes (15)-(17) 
and those of Ref. \cite{universal}
reflects the fact 
that the Dirac equation (1)  
behaves semi-classically
in the limit of a very large orbital
angular momentum.

Let us turn now to find the Regge solutions.
From Eqs. (8) and (9),
Eqs. (6) and (7)
can be written as,

\begin{equation}
\frac{dg}{dr}\,=-\,r^{2\kappa}\, \biggl[
l\, +\,\bigl(
\kappa_s\,-\,\kappa_v
\bigr)
\biggr]\,f,
\end{equation}

\begin{equation}
\frac{df}{dr}\,=\, r^{-2\kappa}\,\biggl[
l\,
-\,\bigl(
\kappa_s\,+\,\kappa_v
\bigr)
\biggr]\,g,
\end{equation}

\noindent where we have assumed 
$E\,-U\,\simeq\,\frac{l\,+\,\xi}{r}\,\simeq\,\frac{l}{r}$,
$\,\,\,r^{2 \kappa\pm 1}\,\simeq\,r^{2\kappa}$,
and 
\begin{equation}
f\,\equiv\,r^{-\kappa}\,F;
\,\,\,\,\,\,\,\,\,\,\,\,\,\,\,\,\,
g\,\equiv\,r^{\kappa}\,G.
\end{equation}

By substituting (18) in (19) it is obtained

\begin{equation}
g^{\prime \prime}\,-\,2\,\kappa\,g^{\prime}\,+
\,\biggl[(l\,-\,\kappa_v)^2\,-\,\kappa_s^2\biggr]\,g\,=\,0.
\end{equation}

\noindent By solving this equation 
and substituting the result in  
(18), we obtain the Regge solutions 
for a $(Q,\bar{q})$ system

\begin{equation}
G\,=\,r^{-\kappa}\,e^{\kappa\,r}\,
\biggl[
A_1\,e^{\sqrt{\kappa_s^2\,+\,2\,l\,\kappa_v\,-\,\kappa_v^2}\,\,r}
\,+\,
A_2\,e^{-\sqrt{\kappa_s^2\,+\,2\,l\,\kappa_v\,-\,\kappa_v^2}\,\,r}
\biggr],
\end{equation}

\begin{eqnarray}
\begin{array}{lll}
F=-\frac{r^{-\kappa}\,e^{\kappa\,r}}
{(l\,-\kappa_v)\,+\,\kappa_s}\, \times 
\\  
\\
\biggl[
{A_1\,
\biggl(
\kappa\,+\,\sqrt{\kappa_s^2\,+\,2\,l\,\kappa_v\,-\,\kappa_v^2 }}
\biggr)
e^{\sqrt {\kappa_s^2\,+\,2\,l\,\kappa_v\,-\,\kappa_v^2}\,\,r}
+
\\
\\  
A_2\,
\biggl(
\kappa\,-\,\sqrt{\kappa_s^2\,+\,2\,l\,\kappa_v\,-\,\kappa_v^2}
\biggr)
e^{-\sqrt{\kappa_s^2\,+\,2\,l\,\kappa_v\,-\,\kappa_v^2 }\,\,r}
\biggr].
\end{array}
\end{eqnarray}

As can be seen from the above equations a
sinusoidal behavior in the quantity in brackets
implies that $\vert\psi\vert^2$ behaves as
$r^{-k}\,e^{\kappa\,r}$ which means
that $\psi$ is not normalizable
for $\kappa\,=\,l$.
So in what follows we assume
$\kappa_s^2\,+\,2\,l\,\kappa_v\,-\,\kappa_v^2\,>\,0$.
From this assumption it
follows immediately that
if we want 
physically admissible Regge solutions 
({\it i.e.} $\lim_{r \to \infty}\,\psi\,\to \,0$) 
it is necessary to drop also the term 
proportional to $A_1$. Consequently,
in all of the discussions below 
we are taking $A_1\,=\,0$.
Another observation 
concerning Eq. (23) is 
that the strength of the potentials 
must be such that 
$l\,+\,\kappa_s\,\ne \,\kappa_v$
in order to avoid an unphysical
divergence 
in the lower 
component of the solution.

Let us consider first the 
situation where the confinement potential
is strictly scalar ($\kappa_v\,=\,0$).
In this case the solutions are

\begin{equation}
G\,=\,r^{-\kappa}\,
e^{(\kappa\,-\,\kappa_s\,)\,\,r},
\end{equation}

\begin{eqnarray}
\begin{array}{lll}
F&=&-
\,\,\frac{\kappa\,-\,\kappa_s}{l\,\,+\,\kappa_s}
\,\,r^{-\kappa}\,\,e^{(\kappa\,-\,\kappa_s)\,\,r}.
\end{array}
\end{eqnarray}

\noindent As we may observe from these equations, 
if the strength of the scalar potential is
strong enough to compensate the intense  
`centrifugal forces' it is possible 
to find physically admissible solutions
for $r\,\rightarrow\,\infty$ 
for either $\kappa\,=\,l$ and $\kappa\,=\,-l$.

Let us turn now
to consider the case when
the confining potential is exclusively  
vectorial ($\kappa_s\,=\,0$).
As can be seen from Eqs. (22) and (23)
in this case the solutions are 

\begin{equation}
G\,=\,r^{-\kappa}\,
\,e^{[\kappa\,-\,\sqrt{\,(2\,l/\kappa_v\,-\,1)} \,\,\,\kappa_v]\,r},
\end{equation}

\begin{eqnarray}
\begin{array}{lll}
F=-\frac{ \kappa\,-\,\kappa_v\,\sqrt{\,2\,l/\kappa_v\,-\,1\,\,}}
{l\,-\,\kappa_v}\,\,\,r^{-\kappa}\,\,
e^{[\kappa\,-\,\sqrt{\,(\,2\,l/\kappa_v\,-\,1)}\,\,\kappa_v]\,r}.
\end{array}
\end{eqnarray}

\noindent We note from these equation
that the condition needed to avoid 
that the exponentials 
in (22) and (23) become oscillatory is 
that
$\kappa_v\,<\,2\,l$ 
which restricts the value 
of $\kappa_v$. 
On the other hand, if we ask for values 
of $\kappa_v$  in the Regge region, 
$\kappa_v\,\sim\,l$
to compensate for the
intense `centrifugal force', then
the lower component of the wave function
$\psi$ would 
diverge strongly.  
Indeed, if we take 
the limit $\kappa_v \to l$
in Eq. (27) we find

\begin{eqnarray}
\begin{array}{lll}
\lim_{\kappa_v \rightarrow l}
\,F\, 
=\left\{ \begin{array}{ll}
l\,r^{-l}\,\,e^{2l\,r} & \kappa = l 
\\
\\
l\,r^{l}\,e^{-\,2\,l\,r}
\lim_{\kappa_v \to l}\,\frac{1}{l\,-\,\kappa_v} & 
\kappa =-l
\end{array}
\right. 
\end{array}
\end{eqnarray}

\noindent From Eq. (28) we 
conclude that for values of the strength
of the vector potential in the Regge regions
it is not possible to 
find physically admisible solutions.
In the case of a very weak vector potential 
$\kappa_v\,\ll\,l$,  
Eqs. (26) and (27) yield

\begin{equation}
-\,F\,\sim\,G\,
\sim\,\,r^{-\,\kappa}\,e^{\,\kappa\,r } ,
\end{equation}

\noindent  which shows that 
the solutions also
diverge in this case for $\kappa\,= \,l$. 

Equations (28) and (29)
indicate that the norm of the wavefunctions of 
a $(Q,\bar{q})$ system in high orbital angular momentum states 
confined by a vector potential and 
described by a Dirac equation is not finite.

Let us consider
the behavior
of the Regge solutions
when
both kinds of 
potentials $V$ 
and $S$ contribute to the 
confining of the light degree
of freedom.
Suppose first that 
both potentials compete 
in strength, that is 
$\kappa_s\,=\,\kappa_v$,
then Eqs. (22)
and (23) are

\begin{equation}
G\,=\,r^{-\kappa}\,
\,e^{\bigl(1\,-\,\sqrt{\,\frac{2\,\kappa_v\,}{l}}
\bigr)\,\kappa\,\,r},
\end{equation}

\begin{eqnarray}
\begin{array}{lll}
F=-\frac{\kappa\,-\,\sqrt{\,2\,l\,\kappa_v\,}}
{\,l\,}\,\, 
r^{-\kappa}
\,\,e^{\bigl(1\,-
\,\sqrt{\,\frac{2\,\kappa_v\,}{l}}\bigr)\,\kappa\,\,r}.
\end{array}
\end{eqnarray}

\noindent
From these equations 
it is evident that
for
$\kappa_s\,=\,\kappa_v\,\ll\,l$,
the solutions behave in the same way as
(29). Consequently they
diverge when $\kappa\,=\,l$. While for 
$\kappa_s\,=\,\kappa_v\,\sim\,l$
these become

\begin{equation}
G\,\simeq \,r^{-\kappa},
\end{equation}

\begin{equation}
F\, \simeq \,0,
\end{equation}

\noindent and therefore they 
diverge for $\kappa\,=\,-l$.
If 
both potentials 
are very strong  {\it i.e.}
$\kappa_s\,=\,\kappa_v\,\gg\,2\,l$ the solutions 
(30) and (31) diverge for $\kappa\,=\,-l$
and converge for $\kappa\,=\,l$.

Let us consider now the general situation
where both kind of potentials
do not have the same intensity.  
In this case as can be seen from (22)
and (23), the solutions
are finite only for values of 
$
\sqrt{\kappa_s^2\,+\,2\,l\,\kappa_v\,-\,\kappa_v^2
}
$
large enough to compensate the 
intense centrifugal force. Otherwise
they would diverge for $\kappa\,=\,l$.

Systematic studies of 
the very 
different situation where 
the 
$(Q,\bar{q})$ system is in
states with
small values of the 
orbital angular momentum,
were performed  
in Refs. \cite{olsson}
and \cite{avila}.  
In Ref. \cite{olsson} 
it was shown that the slope of 
the IW function at zero recoil point 
constitutes a sensitive test 
for the nature of the confinement. 
The authors of this work then
calculated this quantity 
according to both approaches:
Dirac equation with scalar
confinement and no-pair equations
with vector confinement
obtaining 
$
\xi^\prime (1)_{D}\,\simeq\,-\,0.90,
$
and
$
\xi^\prime (1)_{n.p.}\,\simeq\,-\,1.20,
$
respectively. From these values 
they found that the last quantity
was in better agreement with
heavy-light data than the first one.
Another
interesting thing 
found in \cite{olsson} was that 
with a scalar confinement in a Salpeter
(no-pair) equation there is not Regge 
behaviour. 
Based on the above findings, the authors of \cite{olsson}
concluded that vector confinement
is favored over scalar confinement
when it is used a Salpeter (no-pair) equation.
On the other hand, 
by using a Dirac equation with scalar confinement 
in \cite{avila} it was calculated 
the slope of the IW function 
at zero recoil point. In \cite{avila}, this quantity 
was first calculated such as it is defined
properly {\it e.g.} in the heavy
quark symmetry limit ($M_Q\rightarrow \infty$)
and the value found  
was the same as that of Ref. \cite{olsson}
e. g. $\xi_D\simeq-0.90$. 
To calculate
the 
relativistic
corrections coming from a $M_Q$ finite
to the slope 
in \cite{avila},  
it was found $
\xi^\prime (1)_{D}\,\simeq\,-\,1.01,
$.  This value for the slope is in 
better agreement with data than
the one calculated 
without relativistic corrections.
Therefore in Ref. \cite{avila}
it was concluded that with a Dirac
hamiltonian with relativistic corrections
describing the $(Q,\bar{q})$ system, 
a scalar confinement is favored 
over vector confinement at low orbital
angular momentum states.

Let us summarize all of the analysis 
done above. We have considered here
a $(Q,\bar{q})$ system described by
a Dirac hamiltonian with Lorentz
scalar $(S)$ and vector $(V)$ 
confinement potentials.
In order to analyze the Regge behavior 
of such a system
we found the respective slopes. It 
was found that their values 
are exactly the same as those
predicted by the general confinement
potential model of Ref.\cite{universal}.
To find the general Regge solutions
of such an excited mesonic system,
it was shown that the solutions 
with scalar confining potential are physically
admissible while those coming from a 
vector potential are not.

As a result 
of all of 
the above discussed we can
conclude in general that if a $(Q,\bar{q})$ system 
is described by a
no-pair equation, scalar confinement is not favored.
However, if we use 
a Dirac equation for describing the hydrogenlike
system $(Q,\bar{q})$,
a scalar confinement is prefered over a vector confinement 
for any value of the orbital angular momentum.

{{\bf Acknowledgement} 
I would like to thank to M. Olsson for
his interesting observations.
I thank to N. A. 
and to N. H. with whom part of this 
work was done. I acknowledge support
from CONACyT grant 3135.}

\end{document}